\documentclass[a4paper]{jpconf}
\usepackage{graphicx}
\begin{document}
\title{Steady-state
nonequilibrium dynamical mean-field theory and the quantum Boltzmann
equation}
\author{J. K. Freericks and V. M. Turkowski}
\address{Department of Physics, 
Georgetown University, Washington, DC 20057 USA}
\ead{freericks@physics.georgetown.edu {\rm and} turk@physics.georgetown.edu}
\begin{abstract}
We derive the formalism for steady state
nonequilibrium dynamical mean-field theory
in a real-time formalism along the Kadanoff-Baym contour. The resulting
equations of motion are first transformed to Wigner coordinates (average
and relative time), and then re-expressed in terms of differential
operators.  Finally, we perform a Fourier transform with respect to the
relative time, and take the first-order limit in the electric field
to produce the
quantum Boltzmann equation for dynamical mean-field theory.  We next discuss
the structure of the equations and their solutions, describing how these
equations reduce to the Drude result in the limit of a constant relaxation
time. We also explicitly demonstrate the equivalence between the Kubo and 
nonequilibrium approaches to linear response. There are a number of 
interesting modifications of the conventional
quantum Boltzmann equation that arise due to the underlying bandstructure
of the lattice.
\end{abstract}
\section{Introduction}
Dynamical mean-field theory was introduced in 1989 by Metzner and 
Vollhardt~\cite{metzner_vollhardt} as a particular limit where the spatial
dimension approaches infinity.  They showed that the self-energy is local
in this case, so the many-body problem simplifies to that of an impurity
in a time-dependent field plus a self-consistency 
relation~\cite{brandt_mielsch}. Since then, nearly all of the classic 
problems of many-body physics have been solved (see~\cite{kotliar_review}
for a review).  The generalization of the formalism to the nonequilibrium
case is straightforward, but involves a number of technical complications.
In this chapter, we describe the steady-state limit, where we start
our system in equilibrium as $t_0\rightarrow -\infty$, then we turn on an
external field at a later time $t_0^\prime$  ($t_0^\prime>t_0$)
and take the limit where $t_0^\prime
\rightarrow -\infty$.  We consider only a spatially uniform (but 
possibly time-dependent) electric field here, and ignore all magnetic-field
effects. The Hamiltonian is written as the sum of two terms: (i) a 
time-independent Hamiltonian $\mathcal{H}$ which describes the
field-independent piece, and (ii) the time-dependent piece 
${\mathcal H}^\prime(t)$ arising from the field dependence.  We adjust the
electron density with a chemical potential, by subtracting the term 
$\mu\mathcal{N}$ from the Hamiltonian. The work presented here
is a natural extension of 
the discussion in Mahan's review article~\cite{mahan} to the case of
dynamical mean-field theory.

\begin{figure}[h!]
\begin{center}
\includegraphics[width=3.0in]{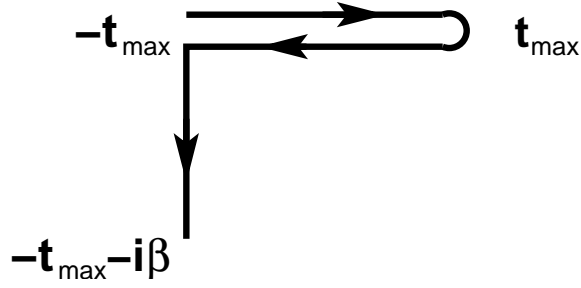}
\end{center}
\caption{\em Kadanoff-Baym
contour for the time variables. The time-domain cutoffs
are symmetric at $\pm t_{\rm max}$ and we take the limit $t_{\rm max}
\rightarrow\infty$.  The arrows indicate the direction to take along the
contour for the time-ordering operation. 
\label{fig: contour}}
\end{figure}

The nonequilibrium formalism works with the so-called contour-ordered Green's
function, which is defined for $t$ and $t^\prime$
on the Kadanoff-Baym contour~\cite{kadanoff_baym}
(see Fig.~\ref{fig: contour}) in terms of the electron creation
($c_{\bf r}^\dagger$) and annihilation ($c^{}_{\bf r}$) operators at
lattice site ${\bf r}$: 
\begin{equation}
G^c_{{\bf r},{\bf r^\prime}}(t,t^\prime)=-i
{\rm Tr}\, {\mathcal T}_c e^{-\beta({\mathcal H}-\mu{\mathcal N})} 
S_c({\mathcal H}^\prime) c_{\bf r}^{}(t)c_{\bf r^\prime}^\dagger(t^\prime)/
\mathcal{Z},
\label{eq: g_contour}
\end{equation}
with $S_c({\mathcal H}^\prime)=\exp[-i\int_c dt^{\prime\prime}
{\mathcal H}^\prime_I(t^{\prime\prime})]$ the evolution operator for the 
time-dependent
piece of the Hamiltonian in the interaction representation, ${\mathcal O}(t)=
\exp[i({\mathcal H}-\mu{\mathcal N})t]{\mathcal O}$
$\exp[-i({\mathcal H}-\mu{\mathcal N})t]$ the time-dependent operator for any
operator ${\mathcal O}$, ${\mathcal Z}={\rm Tr}
\exp[-\beta({\mathcal H}-\mu{\mathcal N})]$, $\beta$ is the inverse temperature,
and ${\mathcal T}_c$ is the time-ordering operator along the Kadanoff-Baym
contour (``earlier'' times ordered along the contour are moved to the
right in the operator expressions). We work in units where $\hbar=c=a=1$
($a$ is the lattice spacing).

The contour-ordered Green's function satisfies the Dyson equation (which
can be thought of as the definition of the contour-ordered self-energy
$\Sigma^c$)
\begin{equation}
G^c_{{\bf r},{\bf r}^\prime}(t,t^\prime)=
G^{c0}_{{\bf r},{\bf r}^\prime}(t,t^\prime)+\int d{\bf r}^{\prime\prime}
\int_c dt^{\prime\prime}\int_c dt^{\prime\prime\prime}
G^{c0}_{{\bf r},{\bf r}^{\prime\prime}}(t,t^{\prime\prime})
\Sigma^c(t^{\prime\prime},
t^{\prime\prime\prime})G^c_{{\bf r}^{\prime\prime},{\bf r}^\prime}
(t^{\prime\prime\prime}
,t^\prime)
\label{eq: dyson_contour}
\end{equation}
with each time
integral being performed over the Kadanoff-Baym contour. The Green's
function $G^{c0}$ is the noniteracting contour-ordered
Green's function in the presence of the
field.  Combining the nonequilibrium perturbation theory
rules described by Langreth~\cite{langreth}
with the perturbation theory expansion of Metzner~\cite{metzner}, shows that the
contour-ordered self-energy is local in infinite dimensions, and hence nonzero 
only when 
${\bf r}={\bf r}^\prime$. This is why the self-energy has no spatial label in
Eq.~(\ref{eq: dyson_contour}). Since our field is spatially uniform, all
Green's functions are translationally invariant, so transforming to momentum
space diagonalizes the spatial part of the Dyson equation (note, we work
in the Hamiltonian or temporal
gauge with a spatially uniform vector potential before making 
the theory gauge-invariant).

Keldysh~\cite{keldysh} was the first to note that if we are
interested in the steady-state limit, as described above, then the
Kadanoff-Baym contour extends from $-\infty$ to $\infty$, and we do not
need to consider the matrix structure associated with the piece of
the branch that extends along the imaginary axis; instead, we take it
into account by incorporating the term $\exp[-\beta({\mathcal H}
-\mu{\mathcal N})]/{\mathcal Z}$ into the operator averages to represent the 
initial equilibrium distribution of the system for times in the distant past.
Next, we can re-express the Dyson equation, which involves an integral
over the Keldysh contour, into an integral over the real-time axis, but at the
price of making the Green's function and self-energy into a $2\times 2$ matrix
with the corresponding Dyson equation~\cite{mahan}:
\begin{eqnarray}
\left ( \begin{array}{cc}
G_{\bf k}^R & G_{\bf k}^K\\
0   & G_{\bf k}^A \end{array}\right )
(t,t^\prime)&=&\left ( \begin{array}{cc}
G_{\bf k}^{R0} & G_{\bf k}^{K0}\\
0   & G_{\bf k}^{A0} \end{array}\right )
(t,t^\prime)\nonumber\\
&+&\int dt^{\prime\prime}\int dt^{\prime\prime\prime}
\left ( \begin{array}{cc}
G_{\bf k}^{R0} & G_{\bf k}^{K0}\\
0   & G_{\bf k}^{A0} \end{array}\right )
(t,t^{\prime\prime})\nonumber\\
&\times&\left ( \begin{array}{cc}
\Sigma^{R} & \Sigma^{K}\\
0   & \Sigma^{A} \end{array}\right )
(t^{\prime\prime},t^{\prime\prime\prime})\times
\left ( \begin{array}{cc}
G_{\bf k}^R & G_{\bf k}^K\\
0   & G_{\bf k}^A \end{array}\right )
(t^{\prime\prime\prime},t^\prime),
\label{eq: dyson2}
\end{eqnarray}
where matrix multiplication is understood amongst the $2\times 2$ matrices
and we have Fourier transformed from real space to momentum space
(the momentum for the Green's functions is {\bf k}).
The three labels (R, K, and A)
on the Green's functions and self-energies refer to
the names retarded, Keldysh, and advanced, respectively; these objects are
formed via specific linear combinations of the contour-ordered Green's
function~\cite{keldysh}. In order to describe these linear combinations, we must
first define four Green's functions from natural choices of the time variables.
The time-ordered Green's function is the 
contour-ordered Green's function with both $t$ and $t^\prime$ on the upper 
real-time branch; the anti-time-ordered Green's 
function has both $t$ and $t^\prime$
on the lower real-time branch; the lesser Green's function has $t$ on
the upper and $t^\prime$ on the lower real-time branches; and the
greater Green's function has $t$ on the lower and $t^\prime$ on the upper 
real-time branches. Then the retarded Green's function is one-half the 
time-ordered minus the anti-time-ordered plus the greater minus the lesser
Green's functions.  The advanced Green's function is one-half the time-ordered
minus the anti-time-ordered minus the greater plus the lesser Green's functions.
The Keldysh Green's function is the greater plus the lesser Green's functions.
There is a conjugate
Dyson equation as well, which looks similar to Eq.~(\ref{eq: dyson2}), but with 
the order of the $G^0\Sigma G$ term switched to $G\Sigma G^0$.

\section{Retarded Green's function}

In this section, we will develop the formalism for the retarded Green's
function.  To start, we examine the equation of
motion (EOM) for the noninteracting retarded Green's function in the presence 
of a uniform electric field.  The field is described with a time-dependent
vector potential ${\bf A}(t)$ and a vanishing scalar potential in the 
Hamiltonian
gauge, where ${\bf E}=-\partial_t {\bf A}(t)$.  This is carried out with
a Peierls substitution~\cite{peierls,jauho_wilkins}: ${\bf k}\rightarrow
{\bf k}-e{\bf A}(t)$ to yield
\begin{eqnarray}
\left [ i\partial_t+\mu-\epsilon_{{\bf k}-e{\bf A}(t)}\right ] G^{R0}_{\bf k}
(t,t^{\prime})&=&\delta(t-t^{\prime}),\nonumber\\
\left [ -i\partial_{t^\prime}+\mu-\epsilon_{{\bf k}-e{\bf A}(t^\prime)}\right ]
G_{\bf k}^{R0}(t,t^\prime)&=&\delta(t-t^\prime),
\label{eq: g_non_eom}
\end{eqnarray}
with $\epsilon_{\bf k}=-t^*\lim_{d\rightarrow\infty}
\sum_{i=1}^d\cos {\bf k}_i/\sqrt{d}$; the hopping matrix element between
nearest neighbors satisfies $t=t^*/2\sqrt{d}$ in the infinite-dimensional
limit~\cite{metzner_vollhardt}.  The next step is to take each 
differential operator in the square brackets in Eq.~(\ref{eq: g_non_eom}) and
operate each on $G^R_{\bf k}(t,t^\prime)$ using the relevant Dyson
equation.  The equation that involves a 
derivative with respect to $t$ uses the Dyson equation in 
Eq.~(\ref{eq: dyson2}),
while the one with respect to $t^\prime$ uses the ``conjugate'' equation.
We then take the sum and the difference of the two equations, and transform
from the time variables $t$ and $t^\prime$ to the Wigner 
coordinates~\cite{wigner} that involve the average and relative times:
$T=(t+t^\prime)/2$ and $t_{\rm rel}=t-t^\prime$. The Green's functions
are now expressed in terms of the Wigner coordinates, for example, 
$G^R_{\bf k}(T,t_{\rm rel})$. Finally, we make the assumption that the
retarded self-energy is a function of the relative time only (which will
be argued to be true
{\it a posteriori} for the Falicov-Kimball model~\cite{falicov_kimball}
below).  The two differential equations become
\begin{eqnarray}
&~&\left [ i\partial_{t_{\rm rel}}+\mu-\frac{1}{2}\epsilon_{{\bf k}+e{\bf E}
(T+t_{\rm rel}/2)}-\frac{1}{2}\epsilon_{{\bf k}+e{\bf E}(T-t_{\rm rel}/2)}
\right ] G^R_{\bf k}(T,t_{\rm rel})=\delta(t_{\rm rel})\nonumber\\
&~&+\frac{1}{2}\int d\bar t
\left [ \Sigma^R(T+\frac{t_{\rm rel}}{2}-\bar t)G^R_{\bf k}
(\frac{T}{2}-\frac{t_{\rm rel}}{4}+\frac{\bar t}{2},\bar t-T+\frac{t_{\rm rel}}
{2})\right.\nonumber\\
&~&+\left.G^R_{\bf k}(\frac{T}{2}+\frac{t_{\rm rel}}{4}+\frac{\bar t}{2},
T+\frac{t_{\rm rel}}{2}-\bar t)\Sigma^R(\bar t-T+\frac{t_{\rm rel}}{2})\right ],
\nonumber\\
&~&\left [ i\partial_T-\epsilon_{{\bf k}+e{\bf E}(T+t_{\rm rel}/2)}+
\epsilon_{{\bf k}+e{\bf E}(T-t_{\rm rel}/2)}\right ] G^R_{\bf k}(T,t_{\rm rel})=
\nonumber\\
&~&\int d\bar t \left [
\Sigma^R(T+\frac{t_{\rm rel}}{2}-\bar t)G^R_{\bf k}(\frac{T}{2}
-\frac{t_{\rm rel}}{4}+\frac{\bar t}{2},\bar t -T+\frac{t_{\rm rel}}{2})
\right.\nonumber\\
&~&\left. -G^R_{\bf k}(\frac{T}{2}+\frac{t_{\rm rel}}{4}+\frac{\bar t}{2},T+
\frac{t_{\rm rel}}{2}-\bar t)\Sigma^R(\bar t-T+\frac{t_{\rm rel}}{2})\right ],
\label{eq: eom2}
\end{eqnarray}
where we have assumed a constant uniform electric field, so ${\bf A}(t)=-{\bf E}
t$; the electric field points along the diagonal $(1,1,1,...)$ of the 
infinite-dimensional hypercube. We need to perform some 
simple manipulations to get these equations into their
final differential form. First we shift 
$\bar t\rightarrow\bar t +T-t_{\rm rel}/2$ in each integral, and then we let 
$\bar t\rightarrow t_{\rm rel}-\bar t$ in the first integral of both equations.
Finally, we use the fact that any infinitely differentiable
function of time $F(T)$ satisfies the Taylor series
operator identity $\exp[\bar t \partial_T]F(T)=F(T+\bar t)$ to produce the 
final time-dependent equations in the Hamiltonian gauge:
\begin{eqnarray}
&~&\left [ i\partial_{t_{\rm rel}}+\mu-\frac{1}{2}
\epsilon_{{\bf k}+e{\bf E}(T+t_{\rm rel}/2)}-\frac{1}{2}
\epsilon_{{\bf k}+e{\bf E}(T-t_{\rm rel}/2)}\right ] G^R_{\bf k}(T,t_{\rm rel})
=\nonumber\\
&~&\delta(t_{\rm rel})
+\int d \bar t \Sigma^R(\bar t)\cosh \left (\frac{\bar t}{2}\partial_T\right )
G^R_{\bf k}(T,t_{\rm rel}-\bar t),\nonumber\\
&~&\left [ i\partial_T-\epsilon_{{\bf k}+e{\bf E}(T+t_{\rm rel}/2)}
+\epsilon_{{\bf k}+e{\bf E}(T-t_{\rm rel}/2)}\right ] G^R_{\bf k}(T,t_{\rm rel})
=\nonumber\\
&~&-2\int d\bar t \Sigma^R(\bar t)\sinh\left (\frac{\bar t}{2}\partial_T\right )
G^R_{\bf k}(T,t_{\rm rel}-\bar t).
\label{eq: eom3}
\end{eqnarray}

The result in Eq.~(\ref{eq: eom3}), is an exact equation of motion satisfied by
the steady-state Green's function.  It is complicated because it has many orders
of the differential operators due to the $\cosh$ and $\sinh$ terms.  It
is also not yet in the form that we can take the linear limit needed for the
quantum Boltzmann equation.  The next step is to introduce a Fourier transform
with respect to the $t_{\rm rel}$ coordinate: 
\begin{equation}
\Sigma^R(t_{\rm rel})=\frac{1}{2\pi}\int d\omega e^{-i\omega t_{\rm rel}}
\Sigma^R(\omega),~G_{\bf k}^R(T,t_{\rm rel})=\frac{1}{2\pi}\int d\omega
e^{-i\omega t_{\rm rel}}G_{\bf k}^R(T,\omega).
\label{eq: fourier}
\end{equation}
When this is substituted in, we replace $\bar t$ terms by $i\partial_\omega$
(this is accomplished by acting the differential operator on the $\exp[-i\omega
t_{\rm rel}]$ term in the Fourier transform and then integrating by parts 
to move the differential operator from the exponential factor onto the 
self-energy factor), and we note that $\cosh(i\alpha)=\cos(\alpha)$ and 
$\sinh(i\alpha)=i\sin(\alpha)$.  Finally, we make the equations gauge invariant.
For our choice of field, this process is straightforward, and we simply
make the coordinate change ${\bf k}\rightarrow
{\bf \bar k}-e{\bf E}T$~\cite{jauho2}.
This requires us to modify the derivative with respect to $T$ via
$\partial_T\rightarrow\partial_T+e{\bf E}\cdot\nabla_{\bf \bar k}$~\cite{mahan}.
Note that
the gauge transformation operation is simply an average-time-dependent shift of
the momentum, so the (local)
self-energy is unchanged.  Furthermore, the gauge-invariant
Green's function $\tilde G^R_{\bf k}(T,\omega)=G^R_{{\bf k}-e{\bf E}T}(T,
\omega)$ turns out to be independent of $T$; we use the tilde symbol to
denote gauge-invariant Green's functions. This last step follows from
the fact that the self-energy has no $T$ dependence, and the (steady-state)
noninteracting
gauge-invariant Green's function (for the simple cosine band), 
expressed as~\cite{davies_wilkins,turkowski}
\begin{equation}
\tilde G^{R0}_{\bf k}(\omega)=\sum_n\frac{J_n\left ( \frac{2\epsilon_{\bf k}}
{eE}\right )}{\omega+\mu-\frac{n}{2}\omega_{\rm Bloch}+i\delta},
\label{eq: g_non_gauge}
\end{equation}
has no $T$ dependence either.  In Eq.~(\ref{eq: g_non_gauge}), $J_n(z)$
is the Bessel function of the first kind with integer order $n$, $E$ is the
magnitude of each component of the electric field along the hypercube
diagonal, and $\omega_{\rm Bloch}=eE$ is the Bloch oscillation frequency; note 
that this choice for the electric field produces the natural Bloch frequency,
even though the norm of the field grows like $\sqrt{d}$.
Since both $\tilde G^{R0}$ and $\Sigma^R$ are independent of $T$, it is a simple
exercise to show that the gauge-invariant version of
Eq.~(\ref{eq: dyson2}), found by making the gauge transformation explicitly
on the equation, leads to a gauge-invariant retarded Green's 
function $\tilde G^R$ that is also independent of $T$. The final EOM becomes
\begin{eqnarray}
&~&(\omega+\mu)\tilde G^R_{\bf k}(\omega)-\frac{\epsilon_{\bf k}}{2}
\tilde G^R_{\bf k}(\omega+\frac{\omega_{\rm Bloch}}{2})-\frac{\epsilon_{\bf k}}
{2}\tilde G^R_{\bf k}(\omega-\frac{\omega_{\rm Bloch}}{2})\nonumber\\
&~&=1+\cos(\frac{1}{2}
\partial_{\omega^\prime}e{\bf E}\cdot\nabla_{\bf k})\Sigma^R(\omega^\prime)
\tilde G^R_{\bf k}(\omega)\Big |_{\omega^\prime=\omega},\nonumber\\
&~&e{\bf E}\cdot\nabla_{\bf k}\tilde G^R_{\bf k}(\omega)-\bar\epsilon_{\bf k}
\tilde G^R_{\bf k}(\omega+\frac{\omega_{\rm Bloch}}{2})+\bar\epsilon_{\bf k}
\tilde G^R_{\bf k}(\omega-\frac{\omega_{\rm Bloch}}{2})\nonumber\\
&~&=2\sin(\frac{1}{2}
\partial_{\omega^\prime}e{\bf E}\cdot\nabla_{\bf k})\Sigma^R(\omega^\prime)
\tilde G^R_{\bf k}(\omega)\Big |_{\omega^\prime=\omega},
\label{eq: eom4}
\end{eqnarray}
with $\bar\epsilon_{\bf k}=-t^*\lim_{d\rightarrow\infty}\sum_{i=1}^d
\sin({\bf k}_i)/\sqrt{d}$.  

The equations in Eq.~(\ref{eq: eom4}) are exact for dynamical mean-field
theory. One can easily verify that the
gauge-invariant noninteracting Green's function in Eq.~(\ref{eq: g_non_gauge})
satisfies the first equation of motion after using the identity
$\alpha J_{n+1}(\alpha)+\alpha J_{n-1}(\alpha)=2n J_n(\alpha)$.  Furthermore,
if the field vanishes, so $E=0$, one can see that the first equation is
satisfied by 
$\tilde G^R_{\bf k}=1/[\omega+\mu-\Sigma^R(\omega)-\epsilon_{\bf k}]$.  The
second equation also holds trivially in equilibrium, because $E=0$. We linearize
Eq.~(\ref{eq: eom4}) in order to find the first nontrivial
parts which are needed for the quantum Boltzmann
equation. The first equation is the same as in equilibrium ($E=0$), with
the same solution~\cite{mahan}, that is, the retarded Green's function is
unchanged by the presence of the field to first order in $E$. The linearized
form of the second equation is
\begin{equation}
e{\bf E}\cdot \left [ \left (1-\frac{\partial\Sigma^R(\omega)}{\partial\omega}
\right ) \nabla_{\bf k}+{\bf v}_{\bf k}\frac{\partial}{\partial\omega}\right ]
\tilde G^R_{\bf k}(\omega)=0,
\label{eq: eom5}
\end{equation}
with ${\bf v}_{\bf k}=\nabla_{\bf k}\epsilon_{\bf k}$ the velocity operator.
It is a simple exercise to show that the equilibrium form of the Green's
function also 
satisfies this equation [just note that $G_{\bf k}^{Req}(\omega)$ depends
on ${\bf k}$ only through $\epsilon_{\bf k}$, so that $\nabla_{\bf k}
G^{Req}_{\bf k}(\omega)={\bf v}_{\bf k}$ $\times
\partial G^{Req}_{\bf k}(\omega)/\partial \epsilon_{\bf k}$].

The final issue that needs to be resolved is that the retarded self-energy 
has no average time dependence.  This is straightforward to show for
the Falicov-Kimball model~\cite{falicov_kimball}, but 
requires too many steps to include the full derivation here.  The basic
idea is that the dynamical mean field for the impurity (which we map onto 
in the infinite-dimensional limit) is average time independent if the
retarded Green's function and initial self-energy are both
average time independent.  Extracting the new self-energy from the impurity
problem in the dynamical mean field does not introduce any average time
dependence because all equations and fields are time-translation invariant
for the steady-state, so the new self-energy is also average time 
independent.  Since
we can start the algorithm with $\Sigma^R=0$, which has no average time 
dependence, the iterative algorithm to solve for the retarded self-energy 
and Green's function never generates any average time-dependence, so those
functions will depend only on $\omega$. It is not clear at this point
whether this holds
for other models, because it is possible that the procedure to determine the
retarded self-energy from the impurity problem in the dynamical mean field
may introduce some average time dependence into the functions.  Such a
complication is straightforward to handle, but for simplicity, we have 
not considered it here.

\section{Keldysh Green's function}

The derivation for the Keldysh Green's function is similar, but it is 
complicated by the fact that the gauge-invariant Keldysh Green's function
may still have average time dependence in the steady state.  This possibility
is related to
the phenomenon of Bloch oscillations~\cite{bloch}, where the steady state
in the absence of scattering is an alternating current, oscillating at
the Bloch frequency, which clearly is a state that has residual time
dependence even in the steady state. This result arises because we are on
a lattice and have restricted ourselves to a single band. In the continuum, 
the Keldysh Green's function is independent of the average time in the steady 
state. On the lattice, if we consider multiple bands, then if there is
tunneling between bands, mediated by the field, then the Bloch oscillations
may also be quenched and the system may evolve into an average time independent
system.

Since we have translational invariance, the spatial part of our problem
diagonalizes when we work in momentum space, which we do here.
We begin with the Dyson equation for the Keldysh Green's function,
which follows from taking the Keldysh component of the full Dyson
equation in Eq.~(\ref{eq: dyson2}).  We can do the same with the ``conjugate''
equation as well.  Finally, we can eliminate the Keldysh Green's function 
on the right-hand side by 
careful manipulation of the retarded or advanced Green's function Dyson 
equations. Hence, there are three forms for the Dyson equation of the Keldysh
Green's function. We write each down schematically, with the notation
that two time-dependent objects adjacent to one another imply a time
integration over the real axis (not the Kadanoff-Baym contour):
\begin{eqnarray}
G^K_{\bf k}&=&G^{K0}_{\bf k}+G^{R0}_{\bf k}\Sigma^RG^K_{\bf k}+G^{R0}_{\bf k}
\Sigma^KG^A_{\bf k}+G^{K0}_{\bf k}\Sigma^AG^A_{\bf k},\nonumber\\
&=&G^{K0}_{\bf k}+G^R_{\bf k}\Sigma^RG^{K0}_{\bf k}+G^R_{\bf k}\Sigma^K
G^{A0}_{\bf k}+G^K_{\bf k}\Sigma^AG^{A0}_{\bf k},\nonumber\\
&=&(1+G^R_{\bf k}\Sigma^R)G^{K0}_{\bf k}(1+\Sigma^AG^A_{\bf k})+G^R_{\bf k}
\Sigma^KG^A_{\bf k}.
\label{eq: dyson_keldysh}
\end{eqnarray}

The noninteracting Keldysh Green's function satisfies a simple differential
equation, because it has no discontinuity when $t=t^\prime$ like the advanced
and retarded Green's functions.  The equations of motion are
\begin{equation}
\left [ i\partial_t+\mu-\epsilon_{{\bf k}+e{\bf E}t}\right ]
G^{K0}_{\bf k}(t,t^\prime)=0,\quad
\left [-i\partial_{t^\prime}+\mu-\epsilon_{{\bf k}+e{\bf E}t^\prime}\right ]
G^{K0}_{\bf k}(t,t^\prime)=0.
\label{eq: keldysh_non_eom}
\end{equation}
Now the derivation proceeds exactly like it did for the retarded Green's
function.  We operate each of the differential operators in brackets onto
the Keldysh Green's function, and use the relevant Dyson equations, and the
EOM of the noninteracting retarded, advanced, and Keldysh Green's functions
to simplify the equations (the advanced Green's
function is equal to the complex conjugate of the retarded Green's
function with the time coordinates 
interchanged).  Then we go to Wigner coordinates and change 
variables in the integrals. The only difference from the derivation for the 
retarded case is that the Keldysh self-energy generically
depends on the average time.  After some long algebra, we arrive at
\begin{eqnarray}
&\,&\left [ i\partial_{t_{\rm rel}}+\mu-\frac{1}{2}\epsilon_{{\bf k}+
e{\bf E}(T+t_{\rm rel}/2)}-\frac{1}{2}
\epsilon_{{\bf k}+e{\bf E}(T-t_{\rm rel}/2)} \right ] 
G^K_{\bf k}(T,t_{\rm rel})=\frac{1}{2}\int d\bar t \nonumber\\
&\,&\left \{ e^{-\frac{\bar t}{2}\partial_T}
G^K_{\bf k}(T,t_{\rm rel}-\bar t) \Sigma^R(\bar t)
+\left [e^{-\frac{\bar t}{2}\partial_T}
G^A_{\bf k}(T,t_{\rm rel}-\bar t)\right ] \left [ e^{\frac{t_{\rm rel}-\bar t}
{2}\partial_T} \Sigma^K(T,\bar t)\right ]\right .\nonumber\\
&\,& +\left [e^{\frac{\bar t}{2}\partial_T}
G^R_{\bf k}(T,t_{\rm rel}-\bar t)\right ] \left [ e^{-\frac{t_{\rm rel}-\bar t}
{2}\partial_T} \Sigma^K(T,\bar t)\right ]
\left.+e^{\frac{\bar t}{2}\partial_T}
G^K_{\bf k}(T,t_{\rm rel}-\bar t) \Sigma^A(\bar t)\right \},\nonumber\\
&\,&\left [ i\partial_T-\epsilon_{{\bf k}+e{\bf E}(T+t_{\rm rel}/2)}+
\epsilon_{{\bf k}+e{\bf E}(T-t_{\rm rel}/2)}\right ] G^K_{\bf k}(T,t_{\rm rel})=
\int d\bar t \nonumber\\ 
&\,&\left \{ e^{-\frac{\bar t}{2}\partial_T}
G^K_{\bf k}(T,t_{\rm rel}-\bar t) 
\Sigma^R(\bar t)\right.
+\left [e^{-\frac{\bar t}{2}\partial_T}
G^A_{\bf k}(T,t_{\rm rel}-\bar t)\right ] \left [ e^{\frac{t_{\rm rel}-\bar t}
{2}\partial_T} \Sigma^K(T,\bar t)\right ]\nonumber\\
&\,& -\left [e^{\frac{\bar t}{2}\partial_T}
G^R_{\bf k}(T,t_{\rm rel}-\bar t)\right ] \left [ e^{-\frac{t_{\rm rel}-\bar t}
{2}\partial_T} \Sigma^K(T,\bar t)\right ]
\left.-e^{\frac{\bar t}{2}\partial_T}
G^K_{\bf k}(T,t_{\rm rel}-\bar t) \Sigma^A(\bar t)\right \},\nonumber\\
&\,&~
\label{eq: keldysh_eom1}
\end{eqnarray}
which are exact equations satisfied by the Keldysh Green's function.

Next, we introduce the Fourier transform of the functions of $\bar t$ and
$t_{\rm rel}-\bar t$, and replace any remaining factors by derivatives 
with respect to the corresponding frequencies, and then integrating by parts
as necessary to move the derivatives off of the exponential terms and onto the
Green's functions or self-energies.  Finally, we perform the gauge-invariance
transformation by shifting the momentum vector with respect to the
average time.  We use the tilde to denote gauge-invariant Green's functions; 
the self-energies are unaffected by the transformation. The final exact 
equations of motion for the Keldysh Green's function are
\begin{eqnarray}
&\,&(\omega+\mu)\tilde G_{\bf k}^K(T,\omega)-\frac{1}{2}\epsilon_{\bf k}
\tilde G_{\bf k}^K(T,\omega+\frac{1}{2}\omega_{\rm Bloch})-\frac{1}{2}
\epsilon_{\bf k} \tilde G_{\bf k}^K(T,\omega-\frac{1}{2}\omega_{\rm Bloch})
=\nonumber\\
&\,&\frac{1}{2}
\left [ e^{\frac{i}{2}\partial_{\omega^\prime}(\partial_T+e{\bf E}\cdot
\nabla_{\bf k})}\tilde G_{\bf k}^K(T,\omega)\Sigma^R(\omega^\prime)
+e^{\frac{i}{2}
(\partial_{\omega^\prime}e{\bf E}\cdot\nabla_{\bf k}-\partial_\omega\partial_T)}
\tilde G_{\bf k}^A(\omega)\Sigma^K(T,\omega^\prime)\right.\nonumber\\
&\,&\left. +e^{-\frac{i}{2}(\partial_{\omega^\prime}e{\bf E}\cdot\nabla_{\bf k}
-\partial_\omega\partial_T)}\tilde G_{\bf k}^R(\omega)\Sigma^K(T,\omega^\prime)
+e^{-\frac{i}{2}\partial_{\omega^\prime}
(\partial_T+e{\bf E}\cdot\nabla_{\bf k})}
\tilde G_{\bf k}^K(T,\omega)\Sigma^A(\omega^\prime)\right ] ,\nonumber\\
&\,&i\partial_T\tilde G_{\bf k}^K(T,\omega)+ie{\bf E}\cdot\nabla_{\bf k}
\tilde G^K_{\bf k}(T,\omega)\nonumber\\
&\,&-i\bar \epsilon_{\bf k}
\tilde G_{\bf k}^K(T,\omega+\frac{1}{2}\omega_{\rm Bloch})
+i\bar \epsilon_{\bf k} 
\tilde G_{\bf k}^K (T,\omega-\frac{1}{2}\omega_{\rm Bloch})=\nonumber\\
&\,& e^{\frac{i}{2}\partial_{\omega^\prime}(\partial_T+e{\bf E}\cdot
\nabla_{\bf k})}\tilde G_{\bf k}^K(T,\omega)\Sigma^R(\omega^\prime)
+e^{\frac{i}{2}
(\partial_{\omega^\prime}e{\bf E}\cdot\nabla_{\bf k}-\partial_\omega\partial_T)}
\tilde G_{\bf k}^A(\omega)\Sigma^K(T,\omega^\prime)\nonumber\\
&\,& -e^{-\frac{i}{2}(\partial_{\omega^\prime}e{\bf E}\cdot\nabla_{\bf k}
-\partial_\omega\partial_T)}\tilde G_{\bf k}^R(\omega)\Sigma^K(T,\omega^\prime)
-e^{-\frac{i}{2}\partial_{\omega^\prime}
(\partial_T+e{\bf E}\cdot\nabla_{\bf k})}
\tilde G_{\bf k}^K(T,\omega)\Sigma^A(\omega^\prime) ,\nonumber\\
&\,&~
\label{eq: keldysh_eom2}
\end{eqnarray}
where, in both cases, we must take the limit $\omega^\prime\rightarrow\omega$.
In Eq.~(\ref{eq: keldysh_eom2}), we used the facts that the gauge-invariant 
retarded and advanced Green's functions and the local retarded and advanced
self-energies are all independent of the average time $T$.

Note that in equilibrium, with $E=0$, the equilibrium forms for the Keldysh
Green's function and self-energy
\begin{equation}
G^{Keq}_{\bf k}(\omega)=-2i[2f(\omega)-1]{\rm Im}G^R_{\bf k}
(\omega),~\Sigma^{Keq}(\omega)=-2i[2f(\omega)-1]
{\rm Im}\Sigma^R(\omega),
\label{eq: keldysh_eq}
\end{equation}
do satisfy Eq.~(\ref{eq: keldysh_eom2}).  Since the EOM for the Keldysh
Green's function is a homogeneous equation, the distribution-function
factor $[2f(\omega)-1]$ comes from the initial conditions
(or equivalently from the bare Keldysh Green's function) and can be viewed
as a boundary condition for the differential equations.

The easiest way to see that the Keldysh Green's function and self-energy may 
generically have average time dependence, is to start with the gauge-invariant
noninteracting Keldysh Green's function in a field (for the simple cosine
band):
\begin{equation}
\tilde G^{K0}_{\bf k}(T,\omega)=2\pi i [2f(\epsilon_{{\bf k}-e{\bf E}T}-\mu)-1]
\sum_n J_n\left ( \frac{2\epsilon_{\bf k}}{eE}\right )\delta \left (\omega+\mu
-\frac{1}{2}eEn\right).
\label{eq: keldysh_non_green}
\end{equation}
The gauge-invariance transformation, which shifts the momentum wavevector,
makes the Bessel-function and delta-function pieces independent of $T$,
but the Fermi-factor piece [$f(\epsilon)=1/\{\exp(\beta\epsilon)+1\}$]
now picks up average time dependence.  [One might be surprised by the fact that 
the noninteracting Keldysh Green's function has the Fermi-Dirac distribution
that depends on $\epsilon_{\bf k}$ in it, since the equilibrium Green's
function described above had the factor $f(\omega)$ in it---this occurs
because when the field vanishes, either form is correct since the 
noninteracting Keldysh Green's function has a $\delta(\omega+\mu-
\epsilon_{\bf k})$ term in it, but when a field is present, the noninteracting 
Keldysh Green's function has an infinite number of $\delta$ functions coming
from the Wannier-Stark ladder, and the initial equilibrium
condition explicitly requires the
Fermi-Dirac distribution as a function of $\epsilon_{\bf k}-\mu$ not $\omega$.]
This $T$-dependence is simple though:
the band structure is a periodic function of momentum, which implies that the
average time dependence must also be periodic in $T$; so the idea of a steady
state that periodically repeats is possible [indeed, one only gets the
correct Bloch oscillations for the noninteracting system if the form in
Eq.~(\ref{eq: keldysh_non_green}) is used]. We need to show that both
the Keldysh self-energy and the gauge-invariant Keldysh Green's functions
are also periodic with the same period. If so, then the $T$-dependence can 
be treated in a Fourier series. It is straightforward to show from the full
Dyson equation, that if the noninteracting Keldysh Green's function is periodic
in $T$, and the Keldysh self-energy is also periodic in $T$, then the full
Keldysh Green's function is periodic in $T$.   Examining the impurity algorithm,
shows that the Keldysh dynamical mean field is periodic in $T$.  If the 
resulting impurity Green's function and self-energy are also periodic in
time, then the impurity solver maintains the periodicity.  This clearly holds
for the Falicov-Kimball model~\cite{falicov_kimball}, and likely holds for 
many other models as well. Note that this analysis says the most general
steady state must be periodic with the Bloch period.  Such functions
also include the constant function, since a constant is periodic with the
Bloch period.

Since the average time dependence is periodic, with a period 
$2\pi/\omega_{\rm Bloch}$, we can expand the Keldysh Green's function and 
self-energy in a Fourier series as
\begin{eqnarray}
\tilde G^K_{\bf k}(T,\omega)&=&\sum_n
e^{-in\omega_{\rm Bloch}T}\tilde G^K_{\bf k}(n,\omega),\nonumber\\
\Sigma^K(T,\omega)&=&\sum_n
e^{-in\omega_{\rm Bloch}T} \Sigma^K(n,\omega).
\label{eq: keldysh_T_fourier}
\end{eqnarray}
Substituting these expansions into Eq.~(\ref{eq: keldysh_eom2}) produces
essentially the same equations except the Keldysh functions depend on
the Fourier index $n$ and frequency $\omega$, and the derivatives with
respect to $T$ are replaced by $\partial_T\rightarrow -in\omega_{\rm Bloch}$.
Finally, we take this equation and linearize it to obtain the final two
expressions for the quantum Boltzman equation in dynamical mean field
theory:
\begin{eqnarray}
&\,&(\omega+\mu-\epsilon_{\bf k})\tilde G_{\bf k}^K(n,\omega)=
\tilde G_{\bf k}^K(n,\omega){\rm Re}\Sigma^R(\omega)+{\rm Re}
\tilde G_{\bf k}^R(\omega) \Sigma^K(n,\omega)\nonumber\\
&\,&+\frac{i}{2}n\omega_{\rm Bloch}\tilde G^K_{\bf k}(n,\omega)\partial_\omega
{\rm Im}\Sigma^R(\omega)-\frac{1}{2}e{\bf E}\cdot\nabla_{\bf k}
\tilde G_{\bf k}^K(n,\omega)\partial_\omega{\rm Im}\Sigma^R(\omega)\nonumber\\
&\,&+\frac{i}{2}n\omega_{\rm Bloch}\partial_\omega{\rm Im}\tilde G_{\bf k}^R
(\omega)\Sigma^K(n,\omega)+\frac{1}{2}e{\bf E}\cdot\nabla_{\bf k}{\rm Im}
\tilde G^R_{\bf k}(\omega)\partial_\omega\Sigma^K(n,\omega),\nonumber\\
&\,&n\omega_{\rm Bloch}\tilde G_{\bf k}^K(n,\omega)+ie{\bf E}\cdot\nabla_{\bf k}
\tilde G_{\bf k}^K(n,\omega)-i\bar\epsilon_{\bf k}
\omega_{\rm Bloch}\partial_\omega\tilde G_{\bf k}^K(n,\omega)=
\nonumber\\
&\,& 2i\tilde G_{\bf k}^K(n,\omega){\rm Im}\Sigma^R(\omega)-2i{\rm Im}
\tilde G_{\bf k}^R(\omega)\Sigma^K(n,\omega)\nonumber\\
&\,&+n\omega_{\rm Bloch}\tilde G_{\bf k}^K(n,\omega)
\partial_\omega{\rm Re}\Sigma^R(\omega)+ie{\bf E}\cdot\nabla_{\bf k}
\tilde G_{\bf k}^K(n,\omega)\partial_\omega{\rm Re}\Sigma^R(\omega)
\nonumber\\
&\,&-n\omega_{\rm Bloch}\partial_\omega{\rm Re}\tilde G_{\bf k}^R(\omega)
\Sigma^K(n,\omega)+ie{\bf E}\cdot\nabla_{\bf k}{\rm Re}
\tilde G_{\bf k}^R(\omega) \partial_\omega\Sigma^K(n,\omega).
\label{eq: keldysh_quantum_boltzmann}
\end{eqnarray}
These equations are obviously quite complex; the second equation is the one most
closely related to the Blotzmann equation, and is the most important one for
determining transport.  

As a first check, however, we evaluate these equations in equilibrium, with
$E=0$; in this case there is no average time dependence, so we are only 
interested in the $n=0$ equations.  Taking the second equation, solving 
for $G^K$, and substituting into
the first equation for the $G^K$ term that appears on the right hand side,
produces the relationship
\begin{equation}
(\omega+\mu-\epsilon_{\bf k})G^K_{\bf k}(\omega)=(\omega+\mu-\epsilon_{\bf k})
\frac{\Sigma^K(\omega)}{[\omega+\mu-{\rm Re}\Sigma^R(\omega)-\epsilon_{\bf k}]^2
+[{\rm Im}\Sigma^R(\omega)]^2},
\label{eq: keldysh_eom_equilib}
\end{equation}
where the factor in the parenthesis cancels on both sides.  After the 
cancellation, we are left with an expression that is consistent with the
relation between the Keldysh self-energy and Green's function in
equilibrium, but it does not explicitly produce the Fermi-Dirac distribution
terms, as discussed above. The quantum Boltzmann equation typically is employed
to obtain a relation between the self-energies and the Keldysh Green's function;
it is expected that the self-energies will be determined by solving some
kind of many-body theory that describes the system of interest (see the next
section).

Because the retarded and advanced Green's functions are unchanged to linear 
order in the field, and because we are only interested in the terms that are
zeroth or first order in $E$ for the quantum Boltzmann equation, we can
replace the full Green's functions and self-energies by their equilibrium values
when they are multiplied by either $E$ or $\omega_{\rm Bloch}$~\cite{mahan}.
This procedure produces a significant simplification of the results.
For example, since the equilibrium Keldysh Green's function and self-energy
are both independent of $T$, only the $n=0$ component survives, so all
$n\omega_{\rm Bloch}$ terms vanish.

We perform the substitution of the equilibrium terms into their respective 
places, and then undergo some signficant algebra to end up at the final
expression for the quantum Boltzmann equations:
\begin{eqnarray}
&\,&(\omega+\mu-\epsilon_{\bf k})\tilde G_{\bf k}^K(n,\omega)=
\tilde G^K_{\bf k}(n,\omega) {\rm Re}\Sigma^{Req}(\omega)+{\rm Re}
G^{Req}_{\bf k}(\omega)\Sigma^K(n,\omega)\nonumber\\
&\,&-4ie{\bf E}\cdot{\bf v_k}\partial_\omega f(\omega)[{\rm Im} G^{Req}_{\bf k}
(\omega)]^2[\omega+\mu-{\rm Re}\Sigma^{Req}(\omega)-\epsilon_{\bf k}]
\delta_{n0},\nonumber\\
&\,&\tilde G^K_{\bf k}(n,\omega)=\frac{\Sigma^K(n,\omega)}{[\omega+\mu-{\rm Re}
\Sigma^{Req}(\omega)-\epsilon_{\bf k}]^2+[{\rm Im}\Sigma^{Req}(\omega)]^2}
\nonumber\\
&\,&
-4ie{\bf E}\cdot{\bf v_k}[{\rm Im}G^{Req}_{\bf k}(\omega)]^2\partial_\omega
f(\omega)\delta_{n0}.
\label{eq: qbe_final}
\end{eqnarray}
Note that the $\tilde G^K_{\bf k}$ and $\Sigma^K$ terms are not the equilibrium
functions, but contain the equilibrium part plus the first-order shift in
$E$; hence the first-order shift in the Keldysh Green's function arises from
two sources---(i) the term coming from the electric-field dependence
of the Keldysh self-energy and (ii) the explicit term given at the end of
the second equation above (when $n=0$).  

The first equation is identical to the second multiplied by 
$[\omega+\mu-{\rm Re}\Sigma^{Req}(\omega)-\epsilon_{\bf k}]$, so there is
only one independent quantum Boltzmann equation. We use the second equation in 
Eq.~(\ref{eq: qbe_final}), because that is the equation most closely related to
the Boltzmann equation (for a discussion of this issue, see~\cite{mahan}). 
Note further, that if the Keldysh self-energy  
has no average time dependence to first-order in $E$, then neither does the 
Keldysh Green's function.

\section{Linear-response conductivity in dynamical mean field theory}

The current operator is determined by the commutator of the polarization
operator with the Hamiltonian, and can be expressed by
\begin{equation}
{\bf J}=e
\sum_{\bf k}\nabla_{\bf k}\epsilon_{\bf k}c^\dagger_{\bf k}c^{}_{\bf k},
\label{eq: current_operator}
\end{equation}
with $c^{\dagger}_{\bf k}$ ($c^{}_{\bf k}$) the electron creation (annihilation)
operators for electrons with momentum ${\bf k}$; this is the current per spin,
since we are considering spinless electrons throughout this manuscript---adding
spin increases the current by a trivial factor of 2. The term 
$\nabla_{\bf k}\epsilon_{\bf k}={\bf v}_{\bf k}$ is the electron velocity.
We can calculate the expectation value of the current operator by employing the
lesser Green's function $G^<=(G^K-G^R+G^A)/2$ as
\begin{equation}
\langle {\bf J}(T)\rangle=-
\frac{i}{2}e\int \frac{d\omega}{2\pi}\sum_{\bf k}{\bf v_k}G^<_{\bf k}(T,\omega).
\label{eq: current_average}
\end{equation}
Using the equilibrium forms for the retarded and advanced Green's functions
(valid through first-order in $E$) and the quantum Boltzmann equation in
Eq.~(\ref{eq: qbe_final}) allows us to solve for the linear-response current
\begin{equation}
\langle {\bf J}(T)\rangle=e^2\sum_{\bf k}({\bf E}\cdot{\bf v_k}){\bf v_k}
\int \frac{d\omega}{2\pi} 
[-\partial_\omega f(\omega)][{\rm Im}G_{\bf k}^{Req}(\omega)]^2,
\label{eq: current_linresp}
\end{equation}
and then the conductivity $\sum_\beta\sigma_{\alpha\beta}{\bf E}_{\beta}=
{\bf J}_{\alpha}$ becomes
\begin{equation}
\sigma_{\alpha\beta}=\delta_{\alpha\beta}e^2\sum_{\bf k}{\bf v_k}_\alpha^2
\int \frac{d\omega}{2\pi}
[-\partial_\omega f(\omega)][{\rm Im}G_{\bf k}^{Req}(\omega)]^2,
\label{eq: conductivity}
\end{equation}
which is the dynamical mean field theory {\it dc} 
conductivity~\cite{jarrell_cond}.  Note that in our derivation, we used the
fact that the Keldysh self-energy is local and has no momentum dependence,
so it vanishes when multiplied by the velocity and integrated over momentum, 
and only the term explicitly
proportional to ${\bf E}$ in Eq.~(\ref{eq: qbe_final})
survives the summation over momentum.  We did not need to assume that there
was no average time dependence to the Keldysh Green's function, but if
there is average time dependence, it arises from the Keldysh self-energy
and cannot have any momentum dependence.  This derivation is much simpler than
that given in Mahan's review article~\cite{mahan} because the vertex
corrections vanish in infinite dimensions~\cite{khurana}, so there is no
Bethe-Salpeter equation to solve for the current; it simply follows 
directly from the quantum Boltzmann equation. Note that the full proof
of the equivalence between the Kubo formula and the quantum Boltzmann
equation was originally proved by Chen and Su~\cite{su}, who found an additional
term neglected by Mahan, that is required for establishing the equivalence.

\section{Recovering the Drude-Sommerfeld model for transport}

The well-known expression for the Drude conductivity is:
\begin{equation}
\sigma_{dc}=\frac{J}{E}=\frac{ne^{2}\tau}{m},
\label{eq: sigma_drude}
\end{equation}
where $n$ is the electron density,
$m$ is the effective electron mass, and $\tau$ is the (constant) relaxation 
time, introduced phenomenologically by Drude. 

Here we concentrate on the conductivity of a strongly
correlated system at finite temperature; 
all scattering is due to the electron-electron interaction.
Determining the {\it dc} conductivity 
in this case is nontrivial (see, for example \cite{Uhrig}).
We analyze the behavior of the {\it dc} conductivity 
in the Falicov-Kimball model for spinless electrons 
on the infinite-dimensional hypercubic lattice. 
The Hamiltonian can be written in the following form~\cite{falicov_kimball}:
\begin{equation}
{\mathcal H}=-\frac{t^*}{2\sqrt{d}}\sum_{\langle ij\rangle}c_{i}^{\dagger}
c^{}_{j} +U\sum_{i}w_ic_{i}^{\dagger}c_{i},
\label{eq: fk_ham}
\end{equation}
where $c_{i}^{\dagger}$ and $c^{}_{i}$ are creation and annihilation
operators for an itinerant electron at site $i$; $t^*=1$ is the
scaled nearest-neighbor hopping~\cite{metzner_vollhardt} that we
use for our energy unit; $w_{i}$  
is the localized electron number operator (equal to $0$ or $1$); and
$U$ is the local Coulomb repulsion between localized and itinerant
electrons. All electrons are spinless in this example.
We also assume that the system is at half-filling, {\it i.e.},
the particle density for the localized and itinerant
electrons is equal  to $1/2$.

In the limit of vanishing electron-electron interaction
on the hypercubic lattice, an arbitrary constant electric field
directed along the lattice diagonal $(1,1,1,...)$,
produces the current $J(T)$ that
can be calculated analytically from Eq.~(\ref{eq: current_average}).
This current displays the expected Bloch oscillations~\cite{bloch}: 
\begin{equation}
J(T)=\frac{e}{4\pi\sqrt{d}}
\sin (\omega_{\rm Bloch}T)
\int d\epsilon 
\frac{-\partial f(\epsilon)}{\partial\epsilon}
\rho (\epsilon) ,
\label{eq: current_bloch}
\end{equation}
where $\rho(\epsilon) =\exp (-\epsilon^{2})/\sqrt{\pi}$ is the 
free-particle density of states,
and the frequency  of the oscillations is exactly the Bloch frequency 
$\omega_{\rm Bloch}$.
It is interesting to discover whether these oscillations 
survive in the presence of electron-electron scattering, 
as $U$ is turned on.
It is difficult to observe this effect experimentally,
since the period of oscillations for realistic fields
is much larger than the typical scattering time, so the electron wavevector
is randomized by scattering before it can undergo a Bloch oscillation.
In the linear-response regime, the free-electron conductivity
grows with the average time, since it is proportional to $ET$.
The Bloch oscillations are absent because the period
(proportional to $1/E$) diverges as $E\rightarrow 0$.

To calculate the expression for the {\it dc} conductivity
from Eq.~(\ref{eq: conductivity}), one needs to find the equilibrium retarded
Green's function for the interacting system. The case of the 
Falicov-Kimball model in the dynamical mean-field theory limit 
can be solved exactly~\cite{brandt_mielsch,review}. This involves solving
the set of the equations for the local Green's function 
$G^{Req}_{\rm loc}(\omega )$, 
the local self-energy $\Sigma^{Req}(\omega )$, and the
dynamical mean-field $\lambda^{Req}(\omega)$:
\begin{equation}
G^{Req}_{\rm loc}(\omega )=\int d\epsilon\rho (\epsilon )
\frac{1}{\omega +\mu -\epsilon -\Sigma^{Req}(\omega )},
\label{eq: hilbert}
\end{equation}
\begin{equation}
\lambda^{Req}(\omega )= \omega+\mu -[G^{Req}_{\rm loc}(\omega )]^{-1}
-\Sigma^{Req}(\omega ),
\label{eq: lambda}
\end{equation}
\begin{equation}
G^{Req}_{\rm loc}(\omega )=\frac{1-w_{1}}{\omega +\mu -\lambda^{Req}(\omega )}
+\frac{w_{1}}{\omega +\mu -U-\lambda^{Req}(\omega )},
\label{eq: imp}
\end{equation}
where $w_{1}$ is the average particle number for the localized electrons.
In the case of half-filling we have $w_{1}=0.5$ and $\mu =0.5U$. 

Once the retarded Green's function is found, the
{\it dc} conductivity can be calculated from Eq.~(\ref{eq: conductivity}) by 
introducing a joint density of states
\begin{equation}
\rho_{2} (\epsilon ,{\bar \epsilon })=\frac{1}{\pi }\exp \left(
-\epsilon^{2}-{\bar \epsilon}^{2} \right).
\label{eq: joint_dos}
\end{equation}
for the two energy
functions $\epsilon_{{\bf k}}$ and ${\bar \epsilon}_{{\bf k}}$
on the infinite-dimensional hypercubic lattice~\cite{Schmidt,turkowski}
and integrating over $\epsilon$ and ${\bar \epsilon}$.

The integration over ${\bar \epsilon}$ in Eq.~(\ref{eq: conductivity}) is 
straightforward and it gives the following general expression for the 
conductivity~\cite{jarrell_cond}:
\begin{equation}
\sigma_{dc} =\sigma_{0}
\int d\omega [-\partial_{\omega}f(\omega )]\tau (\omega ) ,   
\label{eq: Sigmadc}
\end{equation}
where $\sigma_{0}=e^{2}\pi/2d$ and the relaxation time satisfies
\begin{equation}
\tau (\omega) =
\int d\epsilon \rho (\epsilon )
\frac{[{\rm Im}\Sigma^{Req} (\omega )]^{2}/2\pi^{2}}
{\{[\omega +\mu -{\rm Re}\Sigma^{Req} (\omega )-\epsilon ]^{2}
+[{\rm Im}\Sigma^{Req} (\omega )]^{2}\}^{2}}.   
\label{tau}
\end{equation}

It is possible to analyze the expression for $\sigma_{dc}$ 
analytically in the limit of low temperatures~\cite{tau}.
In this case the derivative $-\partial_{\omega}f(\omega )$ is a function
with a sharp maximum around $\omega =0$, and it is enough to approximate
the relaxation time $\tau (\omega )$ by its low-frequency expansion.
In the metallic case when $U<U_{c}=\sqrt{2}$ it is sufficient
to use just the lowest-order expansion $\tau (\omega)=\tau(0)$, yielding
\begin{equation}
\sigma_{dc}=\sigma_{0}\tau(0)
\end{equation}
which is the Drude form.  The relaxation time is determined by
the zero-frequency imaginary part of the retarded self-energy 
${\rm Im}\Sigma^{Req} (0)$, since the real part vanishes
${\rm Re}\Sigma^{Req} (\omega)= (1-Z)\omega\rightarrow 0$ as 
$\omega\rightarrow 0$ (note that $Z<0$ for the Falicov-Kimball model
because it is not a fermi liquid).
In this case, the zero-frequency relaxation time can be evaluated as:
\begin{equation}
\tau (0) \simeq \rho (\mu )
\int d\epsilon 
\frac{[{\rm Im}\Sigma^{Req} (0)]^{2}/2\pi^{2}}
{\{[\mu -\epsilon ]^{2}
+[{\rm Im}\Sigma^{Req} (0)]^{2}\}^{2}}=\frac{\rho (\mu )}{4\pi}
\frac{1}{{\rm Im}\Sigma^{Req} (0)}.
\end{equation}
Therefore, the {\it dc} conductivity is inversely proportional 
to the imaginary part of the retarded self-energy in the metallic phase;
this is the standard Drude picture. 

The situation is more complicated in the insulating case
with $U>\sqrt{2}$. The interacting density of states
$\rho_{int }(\omega )=-(1/\pi )G^{Req}_{\rm loc}(\omega )$ develops a 
pseudogap at $\omega=0$.
The real part of the self-energy is large for small $|\omega|$,
since ${\rm Re}\Sigma^{Req}(\omega)$ has a pole at $\omega =0$. The imaginary
part of the self-energy is small in this case except at $\omega=0$ where
it diverges.  In order to analyze 
the dependence of the {\it dc} conductivity on the model parameters
in this case, we expand the local Green's function
in powers of $1/{\rm Re}\Sigma^{Req}(\omega)$ and ${\rm Im}\Sigma^{Req}(\omega)$
in Eq.~(\ref{eq: hilbert}).
The solution of the system of equations (\ref{eq: hilbert})-(\ref{eq: imp})
for the self-energy to lowest order is~\cite{tau}:
\begin{equation}
{\rm Re}\Sigma^{Req} (\omega)=\frac{U^{2}-2}{4\omega},
\end{equation}
\begin{equation}
{\rm Im}\Sigma^{Req}(\omega)=-\pi\frac{[U^{2}-2]^{3}}{64\omega^{4}}
\rho\left[
\frac{U^{2}-2} {4\omega}
\right]
-\frac{1}{4}\pi[U^{2}-2]\delta (\omega ).
\end{equation}
The lowest-order approximation gives the following expression for the 
relaxation time~\cite{tau}:
\begin{equation}
\tau (\omega )=\frac{16\omega^{4}}{\pi^{2}[U^{2}-2]^{3}}.
\end{equation}
Substituting this expression into Eq.~(\ref{eq: Sigmadc}) yields ($T=1/\beta$
is the temperature here):
\begin{equation}
\sigma_{dc}=\sigma_{0}\frac{T^{4}}{[U^{2}-2]^{3}}
\frac{4}{\pi^{2}}\int_{-\infty}^{\infty}dx\frac{x^{4}}{\cosh^{2}(x)}=
\sigma_{0}\frac{T^{4}}{[U^{2}-2]^{3}}
\frac{7\pi^{2}}{30}.
\end{equation}
The conductivity in the insulating phase goes to zero as $T^{4}$, 
not exponentially, as expected for a conventional insulator with a true gap.
This arises from the fact that there are a small number of states at low
energy, but they have an exceedingly long lifetime, so they can carry a
significant amount of current.

\section{Nonlinear response}

We have derived a series of exact differential equations that are satisfied by
the nonequilibrium Green's functions in the steady state when a uniform
constant electric field is applied [Eqs.~(\ref{eq: eom4}) and 
(\ref{eq: keldysh_eom1})].  These equations appear to be intractable
for a direct solution, because they involve differential operators to infinite
order, and one needs to solve them with alternative strategies.  Instead, we
examined the linear-response regime, and found that we obtain
the exact {\it dc} conductivity for dynamical mean field theory that was
derived independently from the Kubo formula~\cite{jarrell_cond}.  
One can ask, what happens
when we go beyond the linear-response regime and examine nonlinear response?
In the noninteracting case, the system will develop Bloch oscillations, so
the presence of a {\it dc} field creates an {\it ac} current response.
As the scattering is turned on, a classical Boltzmann equation analysis
predicts that the oscillations will survive for
a while, and then ultimately decay due to the scattering.  Is it possible that
in the steady state there are still oscillations, just with a reduced amplitude,
or does the steady state always produce a constant current response?  The
answer to this question lies at the heart of whether or not the Keldysh
Green's functions pick up average time dependence proportional to
${\bf v_k}$ (in a periodic fashion)
for the steady state, or whether they have no average time dependence 
proportional to ${\bf v_k}$ (as in
our linear-response example). There does not appear to be any simple way to
resolve this question, but it seems possible to us that the Bloch oscillations
can survive with a reduced amplitude, until the scattering gets to be too
strong, and the system is forced into a constant, linear-response regime,
due to the extremely short lifetimes.  If this happens, it would be a truly 
quantum effect, since it does not seem to occur in the semiclassical
Boltzmann equation. We intend to work further on this
problem by directly investigating the solution of the Green's functions
and self-energies in the steady state using techniques other than those
derived from the equations of motion that have infinite-order differential
operators in them.

\section{Conclusion}

In this work we have generalized the quantum Boltzmann equation approach of
Mahan~\cite{mahan} to the case of dynamical mean field theory in a single-band
model.  This approach provides us with many simplifications, such as the
fact that the self-energy has no momentum dependence, hence all momentum
derivatives vanish. But it is complicated by some additional nonlinear
effects brought on by the presence of the periodic band structure, which
ultimately relate to the question of whether or not the system has Bloch
oscillations, and how they evolve as the scattering is made stronger.
Since we examined a linear-response limit to derive and use the quantum
Boltzmann equation, we are not able to directly answer such questions here,
instead we will do so in future publications.

We were able to show that taking the linear limit of the nonequilibrium
equations of motion, coupled with a simple theory for scattering by static
charge defects, produced a quantum generalization of the semiclassical
theory for transport with similar results to those of the Drude-Sommerfeld
model. Along the way, we derived a series of exact differential equations
that the nonequilibrium Green's functions satisfy, we showed how to
incorporate periodic average time dependence into this framework if it
exists in the system, and we established an exact equivalence between the
quantum Boltzmann equation approach and the Kubo approach for the 
{\it dc} conductivity.

\ack

We would like to acknowledge useful discussions with J. Serene and V. Zlati\'c.
We acknowledge support of the National Science Foundation (US) under grant
number DMR-0210717 and from the Office of Naval Research (US) under grants
numbered N00014-99-1-0328 and N00014-05-1-0078. We also thank the organizers
of the Workshop for giving us an opportunity to present our results and the
referee for pointing out that a simple Boltzmann equation treatment of 
the response always yields a constant response for the steady state.

\end{document}